\def\ie{{\it i.e.,}\,}

\def\la{\hbox{\raise.5ex\hbox{$<$} 
    \kern-1.1em\lower.5ex\hbox{$\sim$}}} 
\def\ga{\hbox{\raise.5ex\hbox{$>$} 
    \kern-1.1em\lower.5ex\hbox{$\sim$}}}

\newcommand{\Msun}{\mbox{$M_\odot$}}         
\newcommand{\cms}{\mbox{\ cm s${}^{-1}$}}    

\documentclass[graybox, natbib, footinfo]{svmult}

\usepackage{mathptmx}       
\usepackage{helvet}         
\usepackage{courier}        
\usepackage{type1cm}        
\usepackage{makeidx}         
\usepackage{graphicx}        
\usepackage{multicol}        
\usepackage[bottom]{footmisc}

\makeindex             

\begin{document}
\bibliographystyle{plainnat}

\title*{Hydrodynamic simulations of shell convection in stellar cores}
\author{Miroslav Moc\'ak, Ewald M\"uller and Lionel Siess}
\institute{Miroslav Moc\'ak \at Institute of Astronomy and Astrophysics at the Universit\'e Libre de Bruxelles, Belgium \email{mmocak@ulb.ac.be}
\and Ewald M\"uller \at Max-Planck-Institut f{\"u}r Astrophysik, Garching bei M\"unchen, Germany \email{ewald@mpa-garching.mpg.de}
\and Lionel Siess \at Institute of Astronomy and Astrophysics at the Universit\'e Libre de Bruxelles, Belgium \email{siess@astro.ulb.ac.be}}
%
%
\maketitle

\abstract{
Shell convection driven by nuclear burning in a stellar core is a
common hydrodynamic event in the evolution of many types of
stars. We encounter and simulate this convection (i) in the helium
core of a low-mass red giant during core helium flash leading to a
dredge-down of protons across an entropy barrier, (ii) in a
carbon-oxygen core of an intermediate-mass star during core carbon
flash, and (iii) in the oxygen and carbon burning shell above the
silicon-sulfur rich core of a massive star prior to supernova
explosion. Our results, which were obtained with the hydrodynamics 
code HERAKLES, suggest that both entropy gradients and entropy
barriers are less important for stellar structure than commonly
assumed. Our simulations further reveal a new dynamic mixing process
operating below the base of shell convection zones.}

\section{Introduction}
\label{sec:1}

Our knowledge of stellar core convection stems from one-dimensional
stellar evolutionary calculations assuming a hydrostatic stellar
structure and describing dynamic processes (like, e.g., convection) by
local and linear theories \citep{CoxGiuli2008}. This approach is
computationally feasible and predicts observables. However, it
involves some degeneracy due to necessary utilization of
observationally constrained free parameters \citep{Montalban2004},
which may vary during the star's evolution, and from star to star. One
way to check the 1D results is by means of hydrodynamic simulations
which are based on the solution of the Navier-Stokes equations and are
essentially parameter free. This approach is computationally
demanding, and in our case does not yet predict any observables.
Nevertheless, our simulations provide some important insight into
intrinsically multidimensional processes connected to core shell
convection, like convective overshooting, and mixing across entropy
barriers. In particular, they revealed a potentially very important,
new mixing process operative below the base of such convection zones.

\begin{figure}
\centerline{\includegraphics[angle=-90]{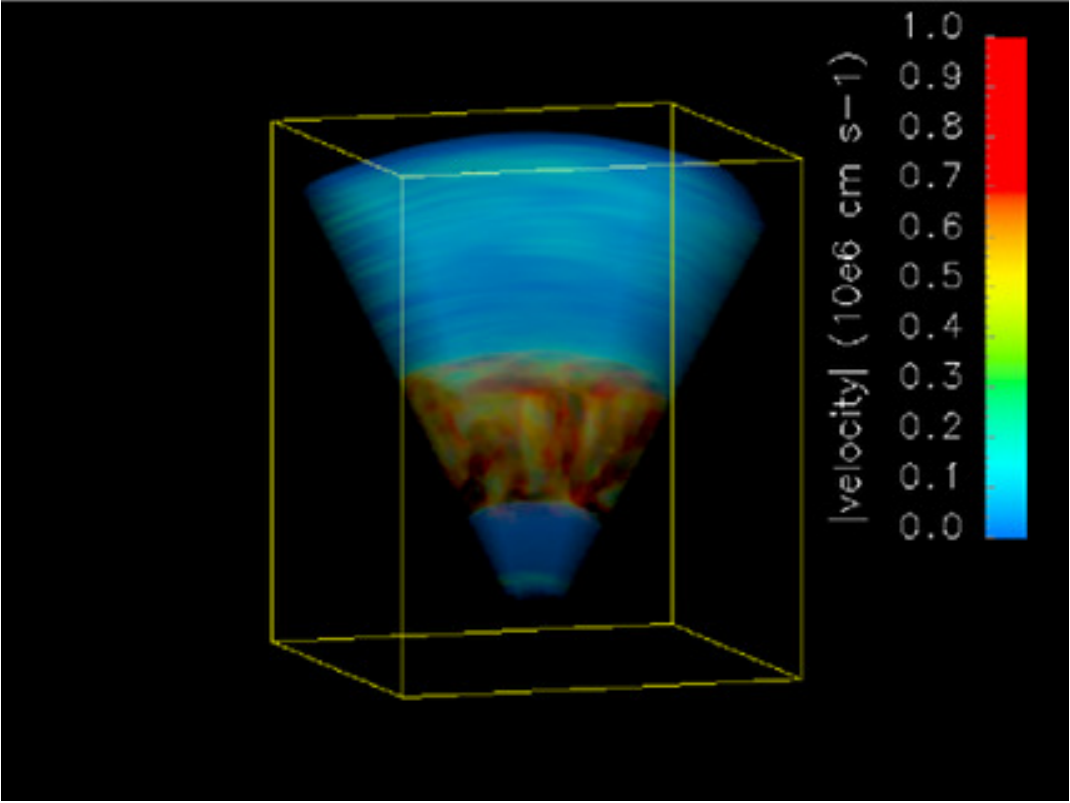}}
\caption{3D volume rendering of the velocity amplitude of shell 
   convection (red/yellow/green) enclosed between two stable layers
   (blue) in the helium core during core helium flash.}
\label{fig:1}       
\end{figure}

\section{Shell convection in stellar cores}
\label{sec:2}

Shell convection in stellar cores occurs in a sandwich-like structure
where a dynamically unstable zone sustained by nuclear burning is
enclosed by two stable regions above and below (Fig.\,\ref{fig:1}).
Besides mixing of chemical species the convective shell redistributes
the energy released by nuclear burning and keeps the stellar core in
quasi-hydrostatic equilibrium. Such convection typically occurs for
instance during:

\begin{enumerate}
\item core helium flashes in low-mass red giants \citep{Mocak2008,Mocak2009,Mocak2010}
\item core carbon flashes of ``super-AGB'' stars \citep{Siess2002} 
\item shell nuclear burning in massive stars \citep{MeakinArnett2007}
\item thermal pulses of AGB stars \citep{Herwig2005,Herwig2006}
\end{enumerate}

\noindent
We simulated the first three cases mostly in two but also in three
spatial dimensions (2D and 3D simulations, respectively) using initial
core structures of metal-rich 1.25\,$\Msun$ \citep{Mocak2008},
9.3\,$\Msun$ star \citep{Siess2006}, and 23\,$\Msun$
\citep{MeakinArnett2007} stars. None of these simulations led to a
violent hydrodynamic event. Initially convection occured in regions
which are unstable according to the Schwarzschild criterion. The
convective velocities found in our 3D simulations are close to those
predicted by mixing-length theory (MLT).  However, our simulations
showed an additional non-radial instability at the boundaries of the
convection zone, which allowed convection to penetrate inexorably into
the adjacent stable layers, thereby increasing the width of the
convection zone on a dynamical timescale. Consequently, core
convection driven by nuclear burning covers likely larger regions than
predicted by MLT theory.

\section{Turbulent entrainment and entropy gradients}
\label{sec:3}

Convection is able to generate mixing in neighboring stable layers by
convective plumes, which can move into these regions due to their
momentum. This process is typically called overshooting and refers to
localized events. However, if the frequency of these events is high,
entropy can change in the affected stable layers. Hence, we can speak
of penetration \citep{Brummell2002}. This occurs due to extensive heat
exchange between the penetrating convective plumes and the stable
layers. Actually, processes at the edges of convection zones appear to
be a combination of both overshooting and penetration. We prefer to
call them turbulent entrainment \citep{Fernando1991} which is well
described by the divergence of the buoyancy flux at convection
boundaries \citep{MeakinArnett2007}. We find turbulent entrainment to
operate at convection boundaries in all our hydrodynamic models. We
thus conclude that convection zones grow in size during dynamic
nuclear flashes (nuclear burning in semi-degenerate gas) when entropy
gradients given by canonical 1D stellar calculations cannot withstand
turbulent entrainment of stable layers. A direct implication of this
result for the core helium flash is the occurence of a hydrogen
injection flash, because the entrainment rates found for the
convection zone driven by helium burning in our models are of the
order of meters per second. Thus, the upper boundary of the convective
shell would reach the overlying hydrogen shell within weeks. This is
not predicted for solar metallicity stars \citep{Campbell2008} as our
initial model. The implications of enlarged convection zones for the
core carbon flash in intermediate-mass stars remain unexplored, and
the case of the oxygen burning shell in massive stars was studied
by \citet{Meakin2006,MeakinArnett2007}.

\section{Hydrogen injection flash and mixing across an entropy barrier}
\label{sec:4}

\begin{figure}
\centerline{\includegraphics[scale=0.33]{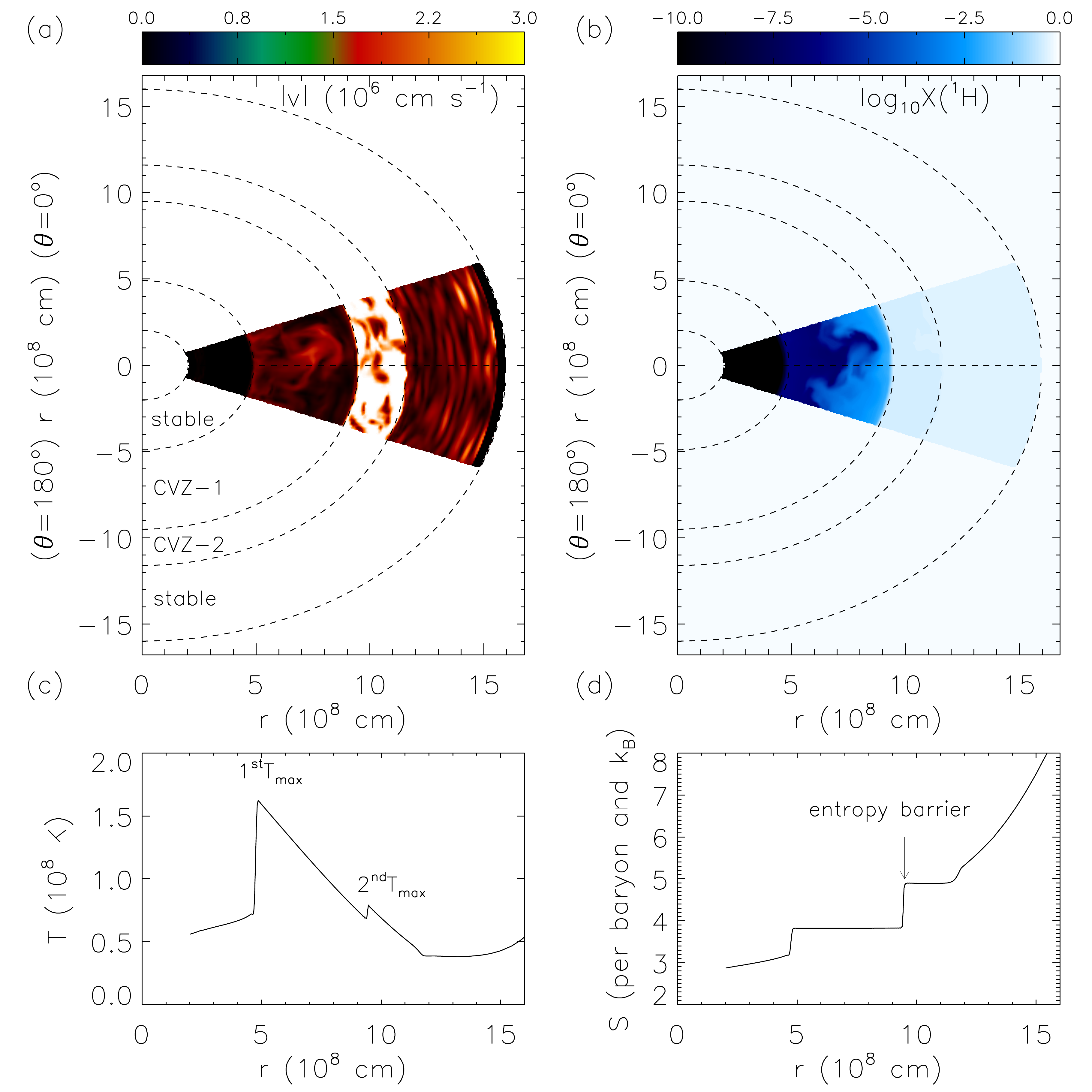}}
\caption{(a) Velocity amplitude $|$v$|$ (in $10^6 \cms$) and (b) 
   hydrogen mass fraction X($^1$H) at the onset of the hydrogen
   injection flash in a meridional plane of a 3D simulation.  Panels
   (c) and (d) show the angle-averaged radial distributions of
   temperature and entropy, respectively, at the same time. The two
   temperature maxima are denoted as 1$^{st}$T$_{max}$ and
   2$^{nd}$T$_{max}$, and the entropy barrier is marked by a vertical
   arrow. Dashed lines enclose distinguished layers, where ``CVZ-1''
   is the convection zone driven by helium burning, ``CVZ-2'' the
   convection zone driven by hydrogen burning, and ``stable'' denotes
   dynamically stable layers.}
\label{fig:2}       
\end{figure}

Hydrogen injection during the core helium flash is predicted by
canonical 1D stellar evolution calculations only for Pop III and
extremely metal-poor stars with intrinsic metallicities [Fe/H] $<
-4$]. A similar hydrogen injection phase also occurs:

\begin{itemize}
\item at the beginning of the thermally pulsing AGB phase of metal-poor
  intermediate-mass stars \citep{Chieffi2001,Siess2002,Iwamoto2004}
\item in ``Late Hot Flasher'' stars experiencing strong mass loss on 
      the RGB \citep{Brown2001,CassisiSchlattl2003}
\end{itemize}

\noindent
We refer to these events as ``dual flashes'' \citep{Campbell2008},
since they all experience simultaneous hydrogen and helium
flashes. These events often lead to a ``splitting'' of the initial
convection zone driven by helium burning due to dredge-down of
material from the above-lying hydrogen-rich envelope and rapid CNO
burning.

In order to study this episode by means of hydrodynamic simulations,
we designed a special stellar model from the helium core of a
metal-rich 1.25$\Msun$, where we shifted the hydrogen-rich layers down
to the upper edge of the already present helium-burning convection
zone. This immediately causes hydrogen to be dredged-down to the
hotter layers of the underlying helium-burning convection zone. The
results of this first 3D simulation of a dual core flash being driven
by hydrogen injection into hot layers of helium-burning convection
zone (Fig.\,\ref{fig:2}) are the following:

\begin{itemize}
\item dredge-down of protons across an entropy barrier between 
      the helium-rich layers and the hydrogen shell 
\item subtle retreat of the helium-burning convection zone (CVZ-1 in 
      Fig.\,\ref{fig:2}) to smaller radii, and the appearance of a
      hydrogen-burning convection zone (CVZ-2 in Fig.\,\ref{fig:2})
      just above it
\item non-existence of stable radiative layer preventing mixing
      between the two convection zones, which is in contradiction to
      1D stellar evolutionary calculations
\item appearance of the $2^{\mbox{nd}}$ temperature peak at the base
      of the hydrogen-burning convection zone, in agreement with 1D
      stellar evolutionary calculations
\end{itemize}

\noindent
We think these results should be qualitatively similar to all dual
flash events and to thermal pulses occurring during
AGB phase of stars. Hydrodynamic simulations of a thermal pulse with
hydrogen mixing into the helium-burning convection zone has already led to
improvements in our understanding of Sakurai's object
\citep{Herwig2011}.

\section{A new dynamic mixing process below the base of shell convection zones}
\label{sec:5}

\begin{figure}
\includegraphics[scale=0.4]{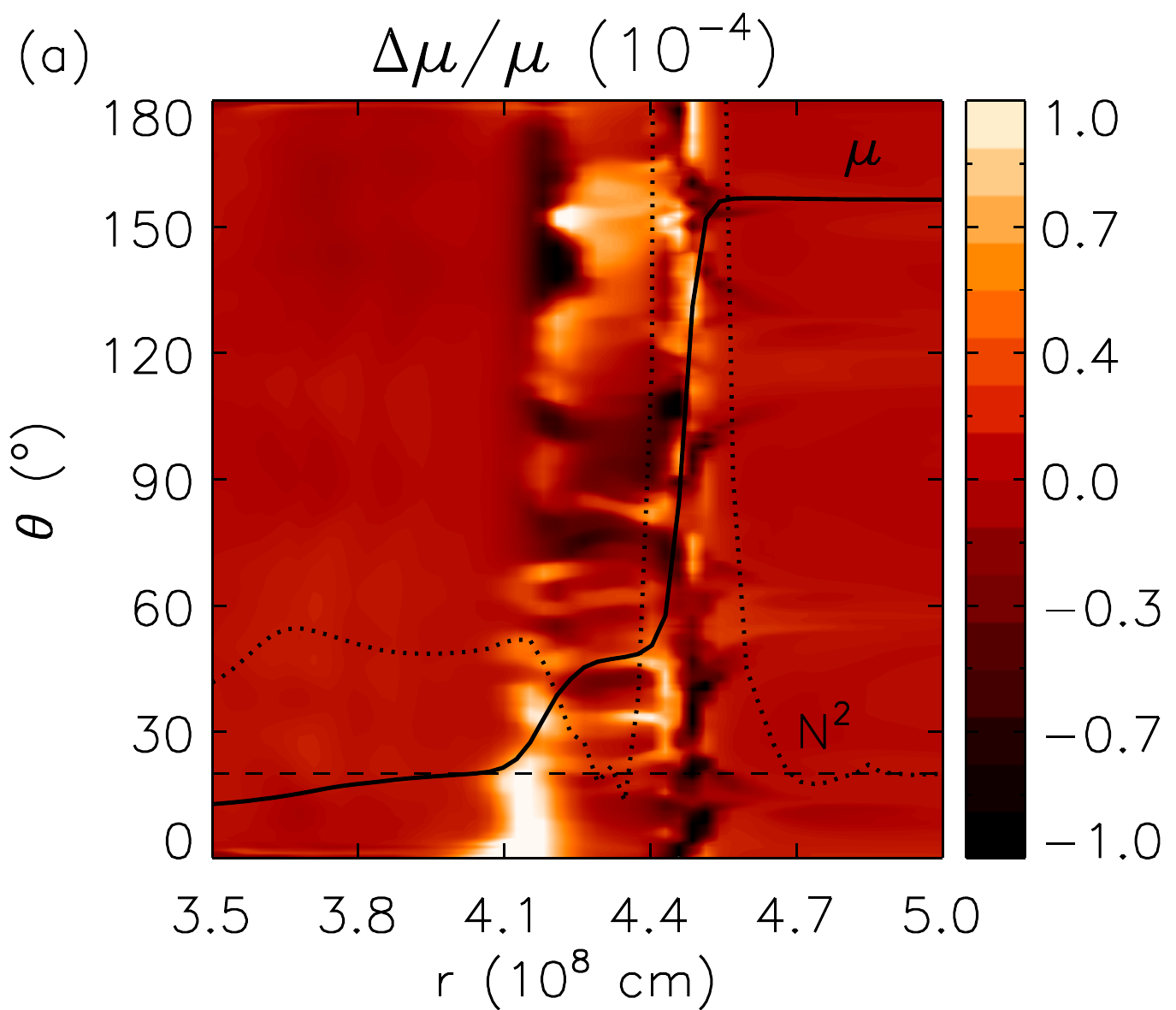}
\includegraphics[scale=0.4]{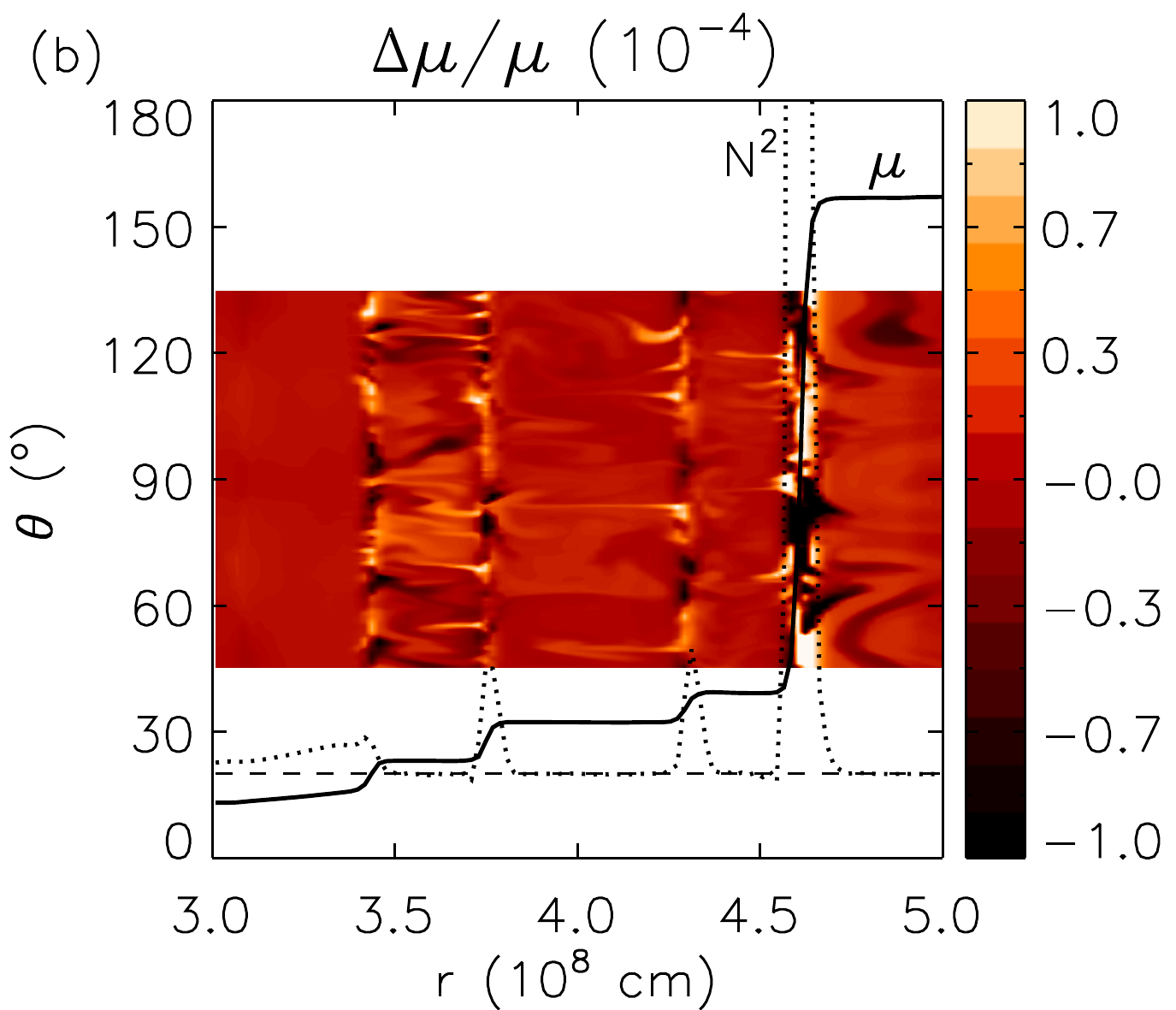}
\centerline{\includegraphics[scale=0.4]{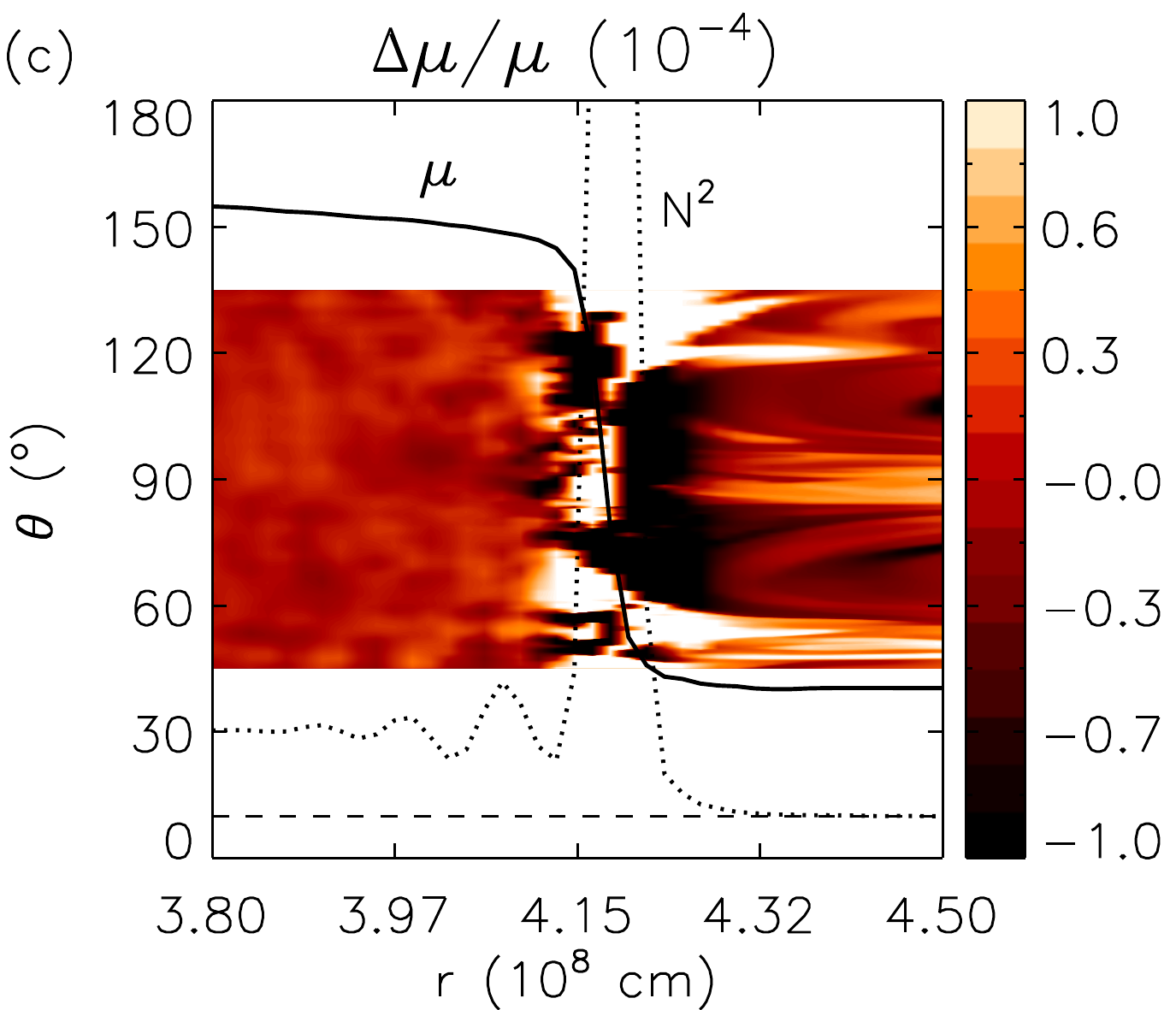}}
\caption{Relative fluctuations of the mean molecular weight defined 
         as $\Delta \mu/\mu = (\mu
         - \langle \mu \rangle_{\theta}) \,/\, \langle \mu \rangle_{\theta}$
         (taken from 2D simulations) below the base of shell
         convection zones during the (a) core helium flash, (b) core
         carbon flash, and (c) oxygen burning shell shown together
         with the radial distributions of the mean molecular weight
         $\mu$ (solid) and the square of the Brunt-V\"ais\"al\"a
         frequency $N^2$. The horizontal dashed line corresponds to
         $N^2 =0$, and $\langle \rangle_{\theta}$ denotes the
         horizontal average at a given radius.}
\label{fig:3}       
\end{figure}

In our 2D and 3D hydrodynamic models of the core helium and carbon
flash, we discovered unreported dynamic mixing process operating at
the base of convection zones. The mixing manifests itself by cold and
over-dense blobs sinking in the direction of gravity, leaving traces
of material with higher mean molecular weight $\mu$
(Fig.\,\ref{fig:3}). Blobs originate from a layer situated just below 
the  lower edge of the convection zone. We do not see any mixing in
our oxygen burning shell models. The driving mechanism of the
mixing remains unclear. Possible explanations for the
appearance of the sinking cold and dense blobs range from:

\begin{itemize}
\item the presence of a steep negative mean molecular gradient 
      $\nabla_{\mu} < 0$ with $\nabla_{\mu} \equiv d \ln \mu / d \ln
      P$ (\ie the molecular weight $\mu$ decreases in direction of
      gravity) destabilizing the layers where the mixing starts
\item a strong shear creating peculiar turbulence at the convection 
      boundary due to trapped gravity waves excited by convection in
      the zone above and indicated by a large value of the square of
      the Brunt-V\"ais\"al\"a frequency N$^2$
\item a weak dynamic stability of the layers below the 
      convection zone indicated by a relatively small positive value
      of N$^2$, which is larger in case of the oxygen burning shell
      model, and hence mixing is effectively suppressed
\item a ``turbulence'' imitating heat transfer or grid scale 
      diffusion removing heat from patches of gas at the convection
      zone boundary, hence making some blobs colder and heavy
\end{itemize}

\section{Summary}
\label{sec:6}

We investigated convection driven by shell nuclear burning during core
helium flash in a low-mass star, core carbon flash in an
intermediate-mass star, and in the oxygen burning shell in a massive
star using data from 2D and 3D hydrodynamic simulations. We find
that all convection zones are growing on a dynamic timescale due to
turbulent entrainment. This either implies that the convection zones
become enlarged or that the physical conditions at the convection zone
boundaries inherited from the initial 1D stellar evolutionary models are
inconsistent with real conditions in stars. The enlargement of the
helium-burning shell convection zone during the core helium flash may
lead to a dual core flash by injection of hydrogen into deeper hot
layers of the helium core. With our specifically designed initial
model for a hydrogen injection flash we show that such an event is
possible and the entropy barrier does not prevent mixing of material
from the hydrogen shell into the underlying hot helium-burning
convection zone. These findings imply that the stabilizing entropy
gradients are less important in stars during flash events. In the
hydrogen injection flash simulation, we observe the appearance of a
second hydrogen-burning shell residing above the already existing
helium-burning convection zone. Radiative stable layers separating the
two convection shells are not observed. We think, that these finding
should be generally valid for all dual flashes. We also found a
new dynamic mixing process below the base of shell convection zones
manifesting itself by cold and dense blobs sinking in direction of
gravity creating eventually finger-like structures. The origin of this
mixing is still under investigation. 

\begin{acknowledgement}
  The simulations were performed at the Rechenzentrum Garching of the Max
  Planck Society. The authors want to thank Casey Meakin for providing us
  with an initial model of the oxygen burning shell and valuable
  discussions. Miroslav Moc\'ak acknowledges financial support from the
  Communaut\'e fran\c caise de Belgique - Actions de Recherche
  Concert\'ees.
\end{acknowledgement}

\bibliography{romeproceedings2010_FINAL}

\end{document}